\documentclass{article}
\usepackage{times}
\usepackage{soul}
\usepackage{url}
\usepackage[hidelinks]{hyperref}
\usepackage[utf8]{inputenc}
\usepackage[small]{caption}
\usepackage{graphicx}
\usepackage{booktabs}
\usepackage{algorithm}

\usepackage{amsthm, amsmath, amssymb, mathtools}
\usepackage{etoolbox}
\usepackage[capitalize]{cleveref}
\usepackage{algorithm}
\usepackage{algorithmicx}
\usepackage[noend]{algpseudocode}
\usepackage{comment}
\usepackage{enumerate}
\usepackage{multicol}
\usepackage[normalem]{ulem}
\usepackage{selectp}
\usepackage{xspace}

\usepackage[format=plain, labelfont=it, textfont=it]{caption}

\theoremstyle{plain}
\newtheorem{theorem}{Theorem}
\newtheorem{corollary}[theorem]{Corollary}
\newtheorem{lemma}[theorem]{Lemma}

\newtheorem{assertion}[theorem]{Assertion}
\newtheorem{observation}[theorem]{Observation}

\theoremstyle{definition}
\newtheorem{definition}[theorem]{Definition}
\newtheorem{example}[theorem]{Example}
\theoremstyle{remark}

\newif\ifcameraready
\camerareadyfalse 

\ifcameraready
   \newcommand{\appwrap}[1]{#1, in the full version \fullversion}
   \outputonly{1-11}
\else
   \newcommand{\appwrap}[1]{#1}
\fi

\renewcommand{\vec}[1]{\bar{#1}}

\newcommand{\FormatRuleClass}[1]{\ensuremath{\mathtt{#1}}\xspace}
\newcommand{\fus}{\FormatRuleClass{fus}}
\newcommand{\bdd}{\FormatRuleClass{bdd}}

\newcommand{\fcs}{\FormatRuleClass{fcs}}

\newcommand{\Exists}[1]{\exists{\scalebox{0.9}{$#1$}}.}
\newcommand{\Forall}[1]{\forall{\scalebox{0.9}{$#1$}}.}

\newcommand{\function}[1]{\mathit{#1}}

\newcommand{\hfun}{\function{h}}
\renewcommand{\hom}{\hfun}

\newcommand{\predicate}[1]{\mathtt{#1}}
\newcommand{\ppred}{\predicate{P}}
\newcommand{\qpred}{\predicate{Q}}
\newcommand{\rpred}{\predicate{R}}

\newcommand{\epred}{\predicate{E}}

\newcommand{\ypred}{\predicate{Y}}

\newcommand{\ginst}{\mathcal{G}}	

\newcommand{\signature}{\mathbb{S}}
\newcommand{\sig}{\signature}
\newcommand{\arity}[1]{\function{ar}(#1)}

\newcommand{\database}{\mathcal{D}}
\newcommand{\db}{\database}
\newcommand{\structure}[1]{\mathcal{#1}}
\newcommand{\instance}{\structure{I}}
\newcommand{\inst}{\instance}
\newcommand{\jnstance}{\structure{J}}
\newcommand{\jnst}{\jnstance}
\newcommand{\adom}[1]{\function{adom}(#1)}

\newcommand{\knowledgebase}{\db, \rs}
\newcommand{\kb}{\knowledgebase}

\newcommand{\set}[1]{\{\,#1\,\}}
\newcommand{\pair}[1]{\langle\, #1 \,\rangle}

\newcommand{\vs}{\vec{s}}

\newcommand{\vx}{\vec{x}}

\newcommand{\vy}{\vec{y}}
\newcommand{\vvv}{\vec{v}}
\newcommand{\vvu}{\vec{u}}

\newcommand{\vz}{\vec{z}}

\newcommand{\vt}{\vec{t}}

\newcommand{\va}{\vec{a}}

\newcommand{\vu}{\vec{u}}

\newcommand{\existentialrule}{\rho}
\newcommand{\eru}{\existentialrule}
\newcommand{\chasesymbol}{Ch}

\newcommand{\chase}[1]{\chasesymbol(#1)}
\newcommand{\step}[2]{\chasesymbol_{#1}(#2)}

\newcounter{rscounter}
\newcounter{dbcounter}
\newcommand{\ruleset}{\mathcal{R}}
\newcommand{\rs}{\ruleset}
\newcommand{\rnf}{\mathtt{cr}(\ruleset)}

\newcommand{\rsp}{\ruleset^{+}}

\newcommand{\dbp}{\database^{+}}

\newcommand{\iffi}{\textit{iff} }

\newcommand{\sh}[1]{\function{sh}(#1)}
\newcommand{\appl}[1]{\function{appl}(#1)}

\newcommand{\rebut}[1]{#1}
\newcommand{\orange}[1]{#1}

\newcommand{\blue}[1]{{\color{blue} #1}}

\newcommand{\nats}{\mathbb{N}}

\newcommand{\cnarule}[3]{#1 \,\to\,\Exists{#2}\; #3}
\newcommand{\narule}[3]{#1 \;\;\to\;\;\Exists{#2}\; #3}

\newcommand{\naruleshort}[3]{#1\to\Exists{#2} #3}

\newcommand{\arule}[3]{#1 &\;\;\to\;\;\Exists{#2}\; #3}
\newcommand{\adrule}[2]{#1 &\;\;\to\;\;#2}
\newcommand{\nadrule}[2]{#1 \;\;\to\;\;#2}

\newcommand{\marking}{\mathbb{m}}

\newcommand{\rpq}{\mathcal{Q}}

\newcommand{\autom}{\mathcal{A}}
\newcommand{\core}{\function{core}}
\newcommand{\states}{\mathbb{Q}}
\newcommand{\state}{\mathtt{q}}
\newcommand{\statef}{\mathtt{q_f}}
\newcommand{\statet}{\mathtt{q_t}}
\newcommand{\inv}[1]{{#1^{-}}}

\newcommand{\rew}{\mathtt{rew}}

\ifcameraready
\usepackage{kr}

\title{\mbox{The Sticky Path to Expressive Querying:} \mbox{Decidability of Navigational Queries under Existential Rules}}

\author{%
Piotr Ostropolski-Nalewaja$^{1,2}$\and
Sebastian Rudolph$^{1,3}$\\
\affiliations
$^1$TU Dresden\\
$^2$University of Wrocław\\
$^3$Center for Scalable Data Analytics and Artificial Intelligence Dresden/Leipzig\\
\emails
postropolski@cs.uni.wroc.pl,
sebastian.rudolph@tu-dresden.de
}
\else
\usepackage{authblk}
\usepackage{natbib}

\title{The Sticky Path to Expressive Querying:\\ Decidability of Navigational Queries under Existential Rules}

\author[1,2]{Piotr Ostropolski-Nalewaja}
\author[1,3]{Sebastian Rudolph}
\affil[1]{TU Dresden}
\affil[2]{University of Wrocław}
\affil[3]{Center for Scalable Data Analytics and Artificial Intelligence Dresden/Leipzig}
\affil[ ]{\textit {postropolski@cs.uni.wroc.pl,
sebastian.rudolph@tu-dresden.de}}
\fi

\begin{document}

\pagenumbering{arabic}
\maketitle

\begin{abstract}

Extensive research in the field of ontology-based query answering has led to the identification of numerous fragments of existential rules (also known as tuple-generating dependencies) that exhibit decidable answering of atomic and conjunctive queries.
Motivated by the increased theoretical and practical interest in navigational queries, this paper considers the question for which of these fragments decidability of querying extends to regular path queries (RPQs). 
In fact, decidability of RPQs has recently been shown to generally hold for the comprehensive family of all fragments that come with the guarantee of universal models being reasonably well-shaped (that is, being of finite cliquewidth).
Yet, for the second major family of fragments, known as finite unification sets (short: fus), which are based on first-order-rewritability, corresponding results have been largely elusive so far.
We complete the picture by showing that RPQ answering over arbitrary fus rulesets is undecidable. 
On the positive side, we establish that the problem is decidable for the prominent fus subclass of sticky rulesets, with the caveat that a very mild extension of the RPQ formalism turns the problem undecidable again.

\end{abstract}

\section{Introduction}

\emph{Existential rules}, also known under the names \emph{tuple-generating dependencies} (TGD)~\cite{AbiteboulHV95}, \emph{Datalog$^+$}~\cite{Gottlob09}, or \emph{$\forall \exists$-rules}~\cite{BagLecMugSal11} have become a very popular formalism in knowledge representation and database theory, with a plethora of applications in ontology-based data access, data exchange, and many more.
One of the fundamental tasks in the context of existential rules is \emph{ontological query answering}, where a query, expressing some information need, is executed over some given data(base) taking into account background knowledge (the \emph{ontology}), which is expressed via a set of existential rules.

The standard query language classically considered in this setting are \emph{conjunctive queries} (CQs) \cite{DBLP:conf/stoc/ChandraM77}, corresponding to the select-project-join fragment of SQL.
As answering\footnote{As \emph{query answering} and \emph{query entailment}, seen as decision problems, are logspace-interreducible, we will use the two terms interchangeably in this paper, sometimes also simply referring to it as \emph{querying}.} of CQs using unconstrained existential rules is undecidable~\cite{ChandraLM81}, much research has been devoted to identifying syntactic and semantic restrictions 
that would warrant decidability of that task. Most investigations in that respect have been along two major lines of research, which can roughly be summarized as \emph{forward-chaining-based} (iteratively applying the rules to the data to see if a query instance is eventually produced) on one side, and \emph{backward-chaining-based} (iteratively applying rules ``backwards'' to the query to see if it is ultimately substantiated by the data) on the other \cite{DBLP:conf/amw/Rudolph14}.  

The advent of NoSQL and graph databases has led to an renewed interest in query languages that overcome some expressivity limitations of plain CQs by accomodating certain forms of recursion \cite{DBLP:conf/pods/RudolphK13}. Among the mildest such extensions are so-called \emph{navigational queries}, which are able to express the existence of size-unbounded structural patterns in the data. Among the most popular such query languages are \emph{regular path queries} (RPQs) and their conjunctive version (CRPQs) as well as their respective 2-way variants (2RPQs, C2RPQs) \cite{DBLP:conf/pods/FlorescuLS98,DBLP:journals/sigmod/CalvaneseGLV03}. In this context, the natural question arises for which of the known classes of rulesets the decidability of CQ entailment generalizes to navigational queries. As it turns out, the situation differs significantly between the forward- and the backward-chaining approaches.

Forward-chaining approaches are based on well-shaped universal models \cite{DBLP:conf/pods/DeutschNR08}, usually obtained via a construction called the \emph{chase} \cite{DBLP:journals/jacm/BeeriV84}.
If one can guarantee the existence of universal models with certain properties (such as being finite or having finite treewidth or -- subsuming the two former cases -- being of finite cliquewidth), CQ entailment is known to be decidable. Fortunately, this generic result was shown to generalize to all query languages which are simultaneously expressible in universal second-order logic and in monadic second-order logic \cite{ICDT2023}, subsuming the all navigational queries considered in this paper.

\begin{corollary}[following from \citeauthor{ICDT2023}, \citeyear{ICDT2023}]
Let $\mathcal{R}$ be a ruleset such that for every database $\mathcal{D}$, there exists a universal model of $\mathcal{D},\mathcal{R}$ having finite cliquewidth. Then C2RPQ answering wrt.~$\mathcal{R}$ is decidable.
\end{corollary}

This very general result establishes in a uniform way decidability of C2RPQ (thus also CRPQ, 2RPQ, and RPQ) answering for all \emph{finite cliquewidth sets} (\fcs) of existential rules, subsuming various classes based on acyclicity \cite{GHKKMMW2013} as well as the guarded family -- including (weakly/jointly/glut-) guarded and \mbox{(weakly/jointly/glut-)} frontier-guarded rulesets \cite{BagLecMugSal11,Cali-GK:guarded-weakly-guarded-tgd-introduced,KrotzschR11}.
For various of these subfragments, decidability of C2RPQ answering had been shown before, partially with tight complexities \cite{ijcai2017-110}.

\medskip

The situation is much less clear (and as of yet essentially unexplored) for backward-chaining-based decidability notions. Rulesets falling into that category are also known as \emph{first-order-rewritable} or \emph{finite unification sets} (\fus). Some types of \fus rulesets simultaneously fall under the forward-chaining case (e.g., linear rules or binary single-head \fus, which are subsumed by \fcs \cite{ICDT2023}) and admit decidable CRPQ entailment on those grounds.
Yet, for \fus rulesets and its popular syntactically defined subclasses such as \emph{sticky rulesets} \cite{sticky-intro}, decidability of path queries has been wide open until now.

In this paper we establish the following results:

\begin{itemize}
\item \orange{Entailment of Boolean RPQs} over \fus rulesets is undecidable (shown by a reduction from the halting problem of Minsky-type two-counter machines).
\item However, the same problem over sticky rulesets is decidable, as will be proven by an elaborate reduction to a finitely RPQ-con\-troll\-able setting.
\item Yet, if we slightly extend \orange{Boolean RPQs} admitting higher-arity pre\-di\-cates in paths, decidability is lost even for sticky rulesets.
\end{itemize}

\section{Preliminaries}
\noindent
\textbf{Structures and homomorphisms.} Let $\mathbb{F}$ be a countably infinite set of {\em function symbols}\orange{, each with an associated arity.} We define the set of {\em terms} $\mathbb{T}$ as a \orange{minimal} set containing three mutually disjoint, countably infinite sets of {\em constants}~$\mathbb{C}$, {\em variables} $\mathbb{V}$, and {\em nulls} $\mathbb{N}$ that \orange{satisfies: $f(\vt) \in \mathbb{T}$ for each tuple $\vt$ of its elements and each symbol $f\in\mathbb{F}$ of matching arity.}
\orange{A {\em signature} $\sig$ is a finite set of predicates.}
We denote the {\em arity} of a predicate
$\ppred$ with $\arity{\ppred}$. An {\em atom} is an expression of the form $\ppred(\vt)$ where $\ppred$ is a predicate and $\vt$ is an $\arity{\ppred}$-tuple of terms. 
Atoms of binary arity will also be referred to as {\em edges}. {\em Facts} are atoms containing only constants. An {\em instance} is a countable (possibly infinite) set of atoms. 
\orange{Moreover, we treat conjunctions of atoms as sets.} A {\em database} is a finite set of facts. The {\em active domain} of an instance $\inst$, denoted $\adom{\inst}$, is the set of terms appearing in the atoms of $\inst$. 
We recall that instances 
naturally represent first-order (\orange{FO}) interpretations.

A {\em homomorphism} from instance $\inst$ to instance $\inst'$ is a function \orange{$\hom: \adom{\inst} \rightarrow \adom{\inst'}$} such that 
%
(1) 
for each atom $\ppred(\vt)$ of $\inst$ we have $\ppred(\hom(\vt)) \in \inst'$, and 
(2) 
for each constant $c \in \mathbb{C}$ we have $\hom(c) = c$.
Given a finite instance $\inst$, a {\em core} of $\inst$ is a minimal subset $\inst'$ such that \orange{$\inst$ homomorphically maps to it.}
It is well known that all cores of a finite structure are isomorphic, allowing us to speak of ``the core'' \orange{-- denoted $\core(\inst)$.}

\smallskip
\noindent
\textbf{Queries.}
A {\em conjunctive query} (CQ) is an FO formula of the 
form: $\Exists{\vx} \phi(\vx, \vy)$ where $\phi$ is a conjunction of atoms over \orange{disjoint tuples of variables} $\vx, \vy$.
The tuple $\vy$ denotes the {\em free variables} of $\phi$. A query with no free variables is {\em Boolean}. 
A {\em union of conjunctive queries} (UCQ) is a disjunction of conjunctive queries having the same tuples of free variables. \orange{Seeing $\phi$ as a set of atoms, the definition of homomorphism naturally extends to functions from CQs to instances}.

A {\em regular path query} (RPQ) \orange{is 
an} expression $\Exists{\vz}A(x,y)$ where $x$ and $y$ are distinct variables, $\vz \subseteq \set{x, y}$, and $A$ is a regular expression over binary predicates from some signature.
Given an instance $\inst$ and two of its terms $s$ and $t$, we write $\inst \models A(s,t)$ to indicate that there exists a directed path $P$ from $s$ to $t$ in $\inst$ whose subsequent edge labels form a word $w$ such that $w$ belongs to the \orange{language of 
$A$.}
Given a signature $\sig$, we define $\inv{\sig}$ as \orange{the 
set} $\{\inv{\sigma} \mid \sigma \in \sig\}$. Given an instance $\inst$ over signature $\sig$, we define 
\orange{an instance} $ud(\inst')$ as $\inst \cup \set{\inv{\ppred}(y,x) \mid \ppred(x,y) \in \inst}$. 
A {\em two-way regular path query} (2RPQ) is \orange{defined as} 
$A(x,y)$ where $x$ and $y$ are variables and $A$ is a regular expression over \orange{predicates}
from $\sig \cup \inv{\sig}$ for signature $\sig$.
\orange{We define C(2)RPQ as a conjunction of (2)RPQs with an existential quantifier prefix.}

\smallskip
\noindent
\textbf{Deterministic Finite Automaton.}
A {\em deterministic finite automaton} (DFA) $\autom$ is a tuple $\pair{\states, \Sigma, \delta, \state_0, \state_{\mathrm{fin}}}$ consisting of a set $\states$ of {\em states}, a finite set $\Sigma$ of symbols called an {\em alphabet}, a {\em transition function} $\delta: \states \times \Sigma \to \states$, a {\em starting} state $\state_0 \in \states$, and an {\em accepting} state $\state_{\mathrm{fin}} \in \states$. 
Let $\delta^{*}: \states \times \Sigma^{*} \to \states$ \orange{ be defined as follows:
$\delta^{*}(\state, \varepsilon) = \state$ if $\varepsilon$ is the empty word, and $\delta^{*}(\state, aw) = \delta^{*}(\delta(\state, a), w)$ if $a \in \Sigma$.}
\orange{$\autom$ {\em accepts}} a word $w \in \Sigma^{*}$ \iffi $\delta^{*}(\state_0, w) = \state_{\mathrm{fin}}$. The {\em language} of $\autom$ is the set of words it accepts. 
Note that for any DFA, 
\orange{one can} construct a regular expression representing the same language, and vice versa.

\subsection{The Chase and Existential Rules}
An FO formula $\rho$ of the form $\Forall{\vx \vy} \alpha(\vx, \vy) \to \Exists{\vz} \beta(\vy, \vz)$ is called an {\em existential rule} (short: rule),
where $\vx, \vy$ and $\vz$ are tuples of variables, $\alpha$ is a conjunction of atoms, and $\beta$ is an atom. We call $\alpha(\vx, \vy)$ the {\em body} of $\rho$ and $\beta(\vy, \vz)$ its {\em head}, while $\vy$ is called the {\em frontier}. CQs $\Exists{\vx} \alpha(\vx, \vy)$ and $\Exists{\vz} \beta(\vy, \vz)$ will be called the {\em body query} and the {\em head query} of $\rho$, respectively. A rule is {\em Datalog} if $\vz$ is empty. 
A finite set of rules is simply called a {\em ruleset}. We may drop the universal quantifier in 
rules 
\orange{for} visual clarity.  
Satisfaction of a rule $\rho$ (a ruleset $\rs$) by an instance $\inst$ is
defined as usual and is written $\inst \models \rho$ ($\inst \models \rs$).
Given a database $\db$ and a ruleset $\rs$, we define an instance $\inst$ to be a {\em model} of $\db$ and $\rs$, written $\inst \models (\db, \rs)$, \iffi $\db \subseteq \inst$ and $\inst \models \rs$.

\smallskip
\noindent
\textbf{\orange{Querying under existential rules}.}
Given a query $Q(\vx)$, a set of rules $\rs$, and a database $\db$ along with a tuple $\va$ of constants, we say that $Q(\va)$ is \emph{entailed} by $\db, \rs$ \iffi every model of $\db$ and $\rs$ satisfies $Q(\va)$.
In such a case we write $\db, \rs \models Q(\va)$ or $\db, \rs, \va \models Q(\vx)$. We then also call the tuple $\va$ a \emph{certain answer} for the query $Q(\vx)$ with respect to $\db, \rs$.
Due to the computational similarity and easy interreducibility of the two tasks, we will not distinguish \emph{query entailment} from \emph{query answering} and use the two terms synonymously in this paper. 

\medskip

Next, we recap a specialized version of the Skolem chase \cite{journey-paper}. While more involved than the ``mainstream'' Skolem chase, we need to employ this variant to establish the required results regarding the various ruleset-transformations presented in the paper. In short, it enforces that the Skolem naming depends only on the shape of the rule head.

\smallskip
\noindent
\textbf{Isomorphism types.} 
Two CQs have the same {\em isomorphism type} if one can be obtained from the other by 
a bijective renaming of its variables (including existentially quantified ones). 
We denote the isomorphism type of a CQ $\Phi$ as $\tau(\Phi)$.
%
For a CQ $\phi(\vy)$ of isomorphism type $\tau$ and any existentially quantified variable $z$ of $\phi(\vy)$, we introduce a $|\vy|$-ary function symbol $f_z^\tau$.

\smallskip
\noindent
\textbf{Skolemization.}
Let $\vz$ be a tuple $\pair{z_1, \ldots z_k}$ of variables.
For a 
CQ $\phi = \Exists{\vz}\psi(\vy, \vz)$ and a mapping $\hom$ \orange{from $\vy$} 
to a set of terms, we define the {\em Skolemization} $\sh{\phi}$ of $\phi$ through $\hom$ as the instance $\psi(\hom(\vy), \pair{z'_1, \ldots z'_k})$  where $z_i' = f_{z_i}^\tau(\hom(\vy))$ and $\tau$ is the isomorphism type of $\Exists{\vz}\psi(\hom(\vy), \vz)$ with $\hom(\vy)$ treated as free variables. We call $\hom(\vy)$ the {\em frontier terms} of $\psi(\hom(\vy), \vz')$.

\smallskip
\noindent
\textbf{\orange{Skolem Chase}.}
Given a ruleset $\rs$ and an instance $\inst$ we call a pair $\pi = \pair{\rho, \hom}$ 
, where $\hom$ is a homomorphism from the body of the rule $\rho \in \rs$ to $\inst$, an \emph{$\rs$-trigger in $\inst$}. We define the {\em application} $\appl{\pi, \inst}$ of the trigger $\pi$ to the instance $\inst$ as $\inst \cup \gamma$ where $\gamma$ is the Skolemization of the head query of $\eru$ through $\hom$. Let $\Pi(\inst, \rs)$ denote the set of $\rs$-triggers in $\inst$.
\newcommand{\kbi}{\inst, \rs}
Given a database $\db$ and a set of rules $\rs$ we define the {\em Skolem chase} $\chase{\kb}$ \orange{as:} 
\begin{align*}
    \step{0}{\kb} &= \db\\
    \step{i+1}{\kb} &= {\small \bigcup}_{\orange{\pi \in \Pi(\step{i}{\db, \rs},\rs)}} \appl{\pi, \step{i}{\db, \rs}}\\
    \chase{\kb} &= {\small \bigcup}_\orange{{i \in \mathbb{N}}} \step{i}{\kb}.
\end{align*}

$\chase{\kb}$ is a universal model for $\rs$ and $\db$, i.e., a model that 
\orange{homomorphically maps} into any 
model \cite{journey-paper}. Thus for any database $\db$, a tuple of its constants $\va$, rule set $\rs$, and a homomorphism-closed query\footnote{That is, a query whose answers are preserved under homomorphisms between instances. Both UCQs and (2)RPQs are known to be homomorphismsclosed. We refer to~\citeauthor{DBLP:conf/pods/DeutschNR08} (\citeyear{DBLP:conf/pods/DeutschNR08}) for a brief discussion.} $\phi(\vx)$ we have: 
$ \chase{\kb} \models \phi(\va) \;  \Longleftrightarrow \;  \kb \models \phi(\va).$ In words: $\phi(\va)$ is entailed by $\kb$ \iffi it is satisfied by the particular instance $\chase{\kb}$.

\noindent
\orange{For an atom  $\alpha \in \chase{\db, \rs}$, we define its \emph{frontier terms} as the terms of $\alpha$ introduced during the chase earlier than $\alpha$.} The \emph{birth atom} of a term $t$ of $\chase{\db, \rs}$ is the atom 
introduced along $t$ during the 
\orange{chase.}
A \emph{join variable} of a rule is a variable appearing more than once in its body.
\orange{We will find it useful to extend the above chase definition in the natural way to not just start from databases, but from arbitrary instance resulting from prior chases -- including infinite ones.}


\subsection{Query Rewritability}

\begin{definition}\label{def:rewriting}
Given a ruleset $\rs$ and a UCQ $Q(\vx)$ we say that a UCQ $Q'(\vx)$ is a {\em rewriting of $Q(\vx)$ (under $\rs$)} if and only if, for every database $\db$ and tuple $\va$ of its constants, we have:
$$\chase{\db, \rs} \models Q(\va) \Leftrightarrow  \db \models Q'(\va).$$
\end{definition}

\smallskip
\noindent
\textbf{Finite Unification Sets} 
A rule set $\rs$ is a {\em finite unification set} (\fus) \iffi every UCQ has a UCQ rewriting under $\rs$. 

\smallskip
\noindent
\textbf{Bounded Derivation Depth Property}
We say 
a ruleset $\rs$ admits the {\em bounded derivation depth property} (is \bdd) \iffi 
for every UCQ $Q(\vx)$ there exists a natural number $k$ such that for every instance $\inst$ and every tuple of its terms $\va$ we have:
$$\chase{\inst, \rs} \models Q(\va) \iff \step{k}{\inst, \rs} \models Q(\va).$$
It turns out that \fus and \bdd classes are equivalent \cite{BDD-FUS}:
\begin{lemma}\label{lem:bdd-is-fus}
A ruleset $\rs$ is \fus if and only if it is \bdd.
\end{lemma}

Rewritings of UCQs can be obtained in a number of ways. For the purpose of this paper, we will rely on the algorithm provided by König et al. (\citeyear{existential-rules-rewriting-procedure-montpelier}). We denote the rewriting of a UCQ $Q$ against a \bdd ruleset $\rs$ obtained through the algorithm presented therein by $\rew(Q, \rs)$, or $\rew(Q)$ in case $\rs$ is known from the context.

\subsection{Sticky Rulesets}

\begin{definition}[Sticky]\label{def:sticky}
Following \citeauthor{sticky-intro} (\citeyear{sticky-intro}),
a ruleset $\rs$ over signature $\sig$ is {\em sticky} \iffi \orange{
there exists a marking $\marking$ of $\sig$ assigning to each $\ppred\in \sig$ a subset of $[1,\arity{\ppred}]$ called {\em marked positions} such that}, for every $\rho \in \rs$,
\begin{itemize}
    \item  if $x$ is a join variable in $\rho$ then $x$ appears at a marked position in the head-atom of $\rho$, and
    \item if $x$ appears in a body-atom of $\rho$ at a marked position then $x$ appears at a marked position in the head-atom of $\rho$.
\end{itemize}
\end{definition}

\begin{observation}\label{obs:terms-stick-in-sticky}
\orange{
Let $\inst$ be an instance, $\rs$ a sticky ruleset, and $t$ a term from $\chase{\inst, \rs}$ with birth atom $\alpha$. If some $t'$ is on a marked position in $\alpha$, then each atom $\beta$ containing $t$ must also contain $t'$ on a marked position.}
\end{observation}
\begin{proof}
We prove this by contradiction.
Let $i$ be the smallest natural number such that there exists an atom $\beta \in \step{i}{\inst, \rs}$ containing $t$ but not $t'$. Take the rule $\rho$ that created $\beta$ and pick some atom $\gamma\in \step{i-1}{\inst, \rs}$ containing $t$ to which one of $\rho$'s body atoms was mapped. Then by assumption $\gamma$ contains $t'$ in a marked position. However, since $\rs$ is sticky, $t'$ must appear in a marked position in $\beta$, leading to a contradiction.
\end{proof}

\begin{definition}\label{def:joinless}
    \orange{Following \citeauthor{Gogacz2017} (\citeyear{Gogacz2017}):} A ruleset is {\em joinless} \iffi none of its rule bodies contains repeated variables.
\orange{Clearly, every joinless ruleset is sticky and therefore is \fus.}
\end{definition}

\subsection{On the Single-Head Assumption}
Note that, throughout this paper,  we assume that all rulesets are ``single-head'', meaning that we disallow conjunctions of multiple atoms in the heads of rules. This is without loss of generality, due to the following transformation presented by \citeauthor{Cali-GK:guarded-weakly-guarded-tgd-introduced}
(\citeyear{Cali-GK:guarded-weakly-guarded-tgd-introduced}):
\newcommand{\rssingle}{\rs_{\mathrm{single}}}
\newcommand{\rsmulti}{\rs_{\mathrm{multi}}}
Given a multi-head ruleset $\rsmulti$ over a signature $\sig$, one can define the ruleset $\rssingle$ as follows:
take a rule $\rho$ of $\rsmulti$
\begin{alignat*}{3}
\arule{\beta(\vx,\vy)}{\vz}{\alpha_1(\vy, \vz), \ldots \alpha_n(\vy, \vz)}    
\end{alignat*}
and replace it by the following collection of rules:
\begin{align*}&\set{\narule{\beta(\vx,\vy)}{\vz}{\ppred_{\rho}(\vy, \vz)}}\\ & \cup\; \set{\nadrule{\ppred_{\rho}(\vy,\vz)}{\alpha_i(\vy, \vz) \; \mid \; i \in [n]}  
}\end{align*}
%
where $\ppred_{\rho}$ is a fresh symbol, unique for each rule $\rho$.

One can note that the chases of both rulesets (over any instance) coincide when restricted to $\sig$, which means the transformation preserves $\fus$ness when restricted to queries over $\sig$. It is also straightforward to note that stickiness is preserved as well – the marking of positions of fresh $\ppred_{\rho}$ symbols can be trivially reconstructed from the markings of the position of the $\alpha_{i}$.

Therefore, all results presented in the remainder of the paper hold for multi-head rules as well.

\section{Undecidability of RPQs over FUS}
\newcommand{\tca}{\mathcal{M}}
\newcommand{\rpqm}{\rpq_{\tca}}
\newcommand{\arpqm}{\autom_\tca}
\newcommand{\cnt}{\mathtt{C}}
\newcommand{\cntl}{\cnt_{\mathtt{x}}}
\newcommand{\cntr}{\cnt_{\mathtt{y}}}
\newcommand{\cntp}[2]{\cnt_{\mathtt{#1}} \texttt{\,:=\;} \cnt_{\mathtt{#1}} \texttt{\,+\,} \mathtt{#2}}
\newcommand{\cntm}[2]{\cnt_{\mathtt{#1}} \texttt{\,:=\;} \cnt_{\mathtt{#1}} \texttt{\,-\,} \mathtt{#2}}
\newcommand{\Deltatt}{\mathtt{d}}
\newcommand{\delt}{\Deltatt}

\begin{theorem}\label{thm:main-undecidable}
\orange{Boolean RPQ} entailment under \fus rulesets is undecidable.
\end{theorem}
We shall prove the above theorem by a reduction from the halting problem for the following kinds of two-counter automata.

\smallskip
\noindent
\textbf{Two-counter automaton.}
A \emph{two-counter automaton (TCA)} consists of a finite set of states $\states$, two positive integer counters called $\cntl$ and $\cntr$, and an instruction $\mathit{instr}(\state) = \pair{\cnt,\delt,\statet,\statef}$ 
for each state $\state \in \states$ which is executed as follows:

\begin{center}
\begin{minipage}{0.80\linewidth}
\begin{algorithmic}[1]
    \If{ $\cnt \texttt{ == } 0$}
    \State $\cntp{}{1}$,\;\; \texttt{move from $\state$ to $\statet$} 
    \Else  
    \State $\cntp{}{\delt}$,\;\; \texttt{move from $\state$ to $\statef$} 
    \EndIf
\end{algorithmic}
\end{minipage}
\end{center}
where $\cnt \in \set{\cntl, \cntr}$ and $\delt \in \set{-1, 1}$.
In each step, the automaton executes the instruction assigned to its current state and moves to the next. We also distinguish a starting state $\state_0 \in \states$ and a halting state $\state_{halt} \in \states$. The \emph{TCA halting problem} asks whether a given TCA $\tca$ can reach the halting state starting from $\state_0$ with both counters set to zero.
\orange{For} a state $\state \in \states$ and 
numbers 
$x$ and $y$ representing the states of counters $\cntl$ and $\cntr$ respectively, we call the tuple $\pair{\state, x, y}$ a {\em configuration} of $\tca$. Note that our TCAs are deterministic.

The \orange{above TCA} is a standard variation of one presented by \citeauthor{book-minsky-machines} (\citeyear{book-minsky-machines}, Chapter 14). The original automaton exhibits two types of instructions: 1) \emph{Add unity to counter $\cnt$ and go to the next instruction}, and 2) \emph{If counter $\cnt$ is not zero, then subtract one and jump to the $n$-th instruction, otherwise go to the next instruction}. It is straightforward to implement Minsky's automata by means of our TCAs. \orange{Thus, we inherit undecidability of the corresponding Halting problem.}

\newcommand{\dbg}{\db_{\mathrm{grid}}}
\newcommand{\rsg}{\rs_{\mathrm{grid}}}

In order to prove \cref{thm:main-undecidable}, we use the following reduction.
Given a TCA $\tca$ we construct a database $\dbg$, ruleset $\rsg$ and a RPQ $\rpq_{\tca}$ such that $\tca$ halts starting from configuration $\pair{\state_0, 0, 0}$ 
 if and only if $\dbg, \rsg \models \rpq_\tca$.

\newcommand{\pnats}{\predicate{Succ}}
\newcommand{\zeropred}{\predicate{Zero}}
\newcommand{\xcoord}{\predicate{XCoord}}
\newcommand{\ycoord}{\predicate{YCoord}}
\newcommand{\yzero}{\predicate{YZero}}
\newcommand{\xzero}{\predicate{XZero}}
\newcommand{\gridpred}{\predicate{GridPoint}}

\subsection{Database $\dbg$ and Ruleset $\rsg$}
Both the database and the ruleset are independent of the specific TCA $\tca$ considered -- thus they are fixed. The idea behind the construction is that $\chase{\dbg, \rsg}$ represents an infinite grid. The database $\dbg$ consists of just the two atoms $\pnats(a, b)$ and  $\zeropred(a)$.

Before stating $\rs_{\mathrm{grid}}$, 
it is convenient to 
define two abbreviations. 
Let $\varphi_\mathrm{right}(x,x',y,z,z')$ denote the CQ
\begin{align*}
    \xcoord(z,\; x) \land \ycoord(z,\; y) &\ \land\\ \xcoord(z', x') \land \ycoord(z', y) &\land \pnats(x,x')
\end{align*}
with the intuitive meaning that \textit{node $z'$ is one to the right from node $z$ on the grid}.
In a similar vein, we let $\varphi_\mathrm{up}(x,y,y',z,z')$ denote the CQ
\begin{align*}
    \xcoord(z,\; x) \land \ycoord(z,\; y) &\ \land\\ \xcoord(z', x) \land \ycoord(z', y')&\land \pnats(y,y')
\end{align*}
meaning that \textit{node $z'$ is one above node $z$ on the grid}.

\newcommand{\incx}{\predicate{IncX}}
\newcommand{\incy}{\predicate{IncY}}
\newcommand{\decx}{\predicate{DecX}}
\newcommand{\decy}{\predicate{DecY}}
\begin{definition}\label{def:undec-ruleset}
We define the grid-building ruleset $\rs_{\mathrm{grid}}$ as the following collection of rules:
\begin{align*}
\arule{\pnats(x,x')}{x''}{\pnats(x',x'')}\\
\arule{\pnats(x,x') \land \pnats(y, y')}{z}{\gridpred(x, y, z)}\\
\adrule{\gridpred(x,y,z)}{\xcoord(z, x)}\\
\adrule{\gridpred(x,y,z)}{\ycoord(z, y)}\\
\adrule{\varphi_\mathrm{right}(x,x',y,z,z')}{\incx(z,z')}\\
\adrule{\varphi_\mathrm{up}(x,y,y',z,z')}{\incy(z,z')}\\
\adrule{\incx(z,z')}{\decx(z',z)}\\
\adrule{\incy(z,z')}{\decy(z',z)}\\
\adrule{\xcoord(z, x) \land \zeropred(x)}{\xzero(z,z)}\\
\adrule{\ycoord(z, y) \land \zeropred(y)}{\yzero(z,z)}
\end{align*}
\end{definition}

\begin{observation}
Ruleset $\rs$ is \fus
\end{observation}
\begin{proof}
Let $\rs_1$ denote the set of the first four rules and let $\rs_2$ denote the rest. Note that: 1) $\rs_1$ is sticky (see \cref{def:sticky}), and thus \fus; 2) $\rs_2$ terminates after a fixed number of chase steps thus is trivially \bdd (and from \cref{lem:bdd-is-fus} it is \fus); and 3) $\chase{\db, \rs} = \chase{\chase{\db, \rs_1}, \rs_2}$.
Then, from the definition of \fus, if $\rs_1$ is \fus and $\rs_2$ is \fus then $\rs$ is \fus{} -- thanks to the ``stratification'' of the two parts of $\rs$, one can obtain the required rewritings by rewriting first against $\rs_2$ and then rewriting the resulting UCQ against $\rs_1$.
\end{proof}

Without going into detail, we note that the observation of $\rs$ being \fus, can also be readily argued using the generic framework for static analysis of rulesets by \citeauthor{BagLecMugSal11} (\citeyear[Section 7]{BagLecMugSal11}), according to which $\rs$ falls into the class ``sticky $\triangleright$ aGRD'' where a ruleset with an acyclic graph of rule depenndencies (in our case: $\rs_2$) is layered ``on top of'' a sticky ruleset (in our case: $\rs_1$). Since both aGRD and sticky are \fus, so is their stratified combination.

\subsection{Grid $\ginst$}

\begin{figure}
\centering
\mbox{\includegraphics[width=\linewidth]{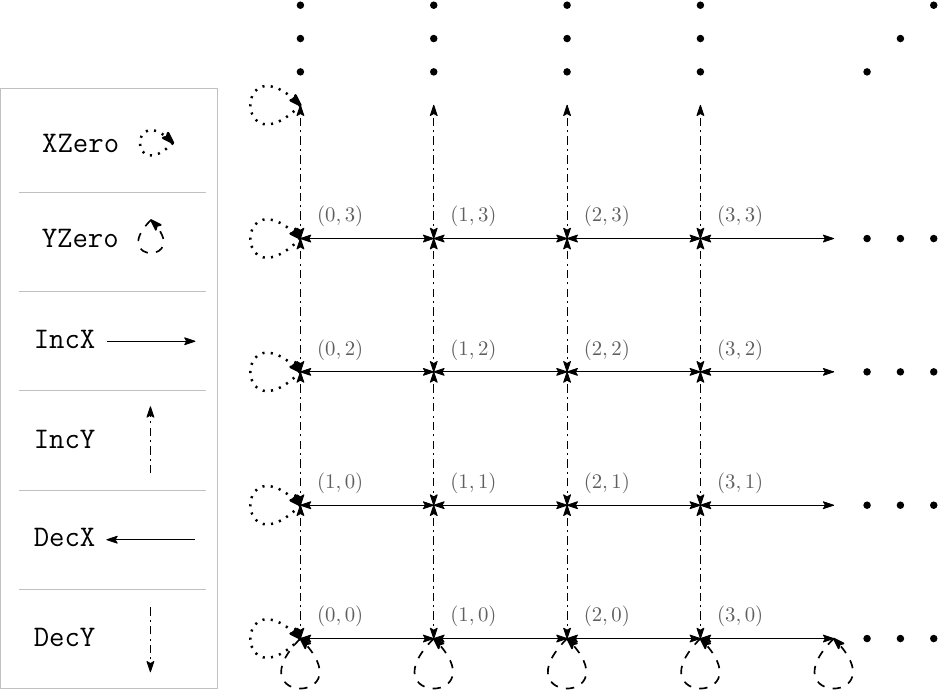}}

\caption{Depiction of $\chase{\dbg, \rsg}$ restricted to predicates $\incx$, $\decx$, $\incy$, $\decy$, $\xzero$, $\yzero$, henceforth denoted $\ginst$.
}\label{fig:grid}
\end{figure}

\Cref{fig:grid} visualizes $\chase{\dbg, \rsg}$ restricted to the \orange{atoms 
using} $\incx$, $\decx$, $\incy$, $\decy$, $\xzero$, and $\yzero$.
As the soon to be defined RPQ $\rpqm$ only uses these binary predicates, it is sufficient to consider this part of $\chase{\dbg, \rsg}$, which \orange{we 
denote} by $\ginst$.

\orange{As $\ginst$ resembles a grid}, we find it convenient to introduce a few notions. First, we identify any term of $\ginst$ with a pair of natural numbers called {\em coordinates}. As usual, the first element of a coordinate pair is called {\em $X$-coordinate}, while the second is called {\em $Y$-coordinate}.

\subsection{RPQ $\rpqm$}

We shall define the Boolean regular path query $\rpqm$ by means of a deterministic finite automaton $\arpqm$.
To simplify its definition, we will allow ourselves to write some expressions which evaluate to binary symbols from $\sig$.%
%
\orange{
We let the function $\Sigma(\cdot)$ assign predicates from $\Sigma$ to operations and tests on the two counters:
\begin{align*}
\Sigma(\cntp{x}{1}) = \incx,\ \ \quad&\quad \Sigma(\cntp{y}{1}) = \incy,\\
\Sigma(\cntm{x}{1}) = \decx,\ \ \quad&\quad \Sigma(\cntm{y}{1}) = \decy,\\
\Sigma(\cntl \texttt{==}\, 0) = \xzero ,\quad&\quad\quad\ \ \ \ \, \Sigma(\cntr \texttt{==}\, 0) = \yzero.
\end{align*}}

\smallskip
\noindent
{\bf  Definition of the automaton $\smash{\arpqm}$. }
The DFA $\smash{\arpqm}$ is a tuple $\smash{\pair{\states_{\autom}, \Sigma_{\autom}, \delta_{\autom}, \state_0, \state_{\mathrm{fin}}}}$ consisting of the set of {\em states} $\states_{\autom}$, the {\em alphabet} $\Sigma_{\autom}$, the {\em transition function} $\delta_{\autom}$, the {\em starting state $\state_0$}, and the {\em accepting state} $\state_{\mathrm{fin}}$. 

\smallskip
\noindent
{\bf States.}
The set $\states_{\autom}$ of states of $\arpqm$ is $\states \cup \states_{aux}$ where $\states_{aux}$ is a set of auxiliary states consisting of four states $\state^{\mathrm{then}_1}$, $\state^{\mathrm{then}_2}$, $\state^{\mathrm{else}_1}$, and $\state^{\mathrm{else}_2}$ for every state $\state \in \states$. The starting (accepting) state of $\arpqm$ is the starting (halting) state of $\tca$.

\smallskip
\noindent
{\bf Alphabet.}
The alphabet $\Sigma_{\autom}$ of $\arpqm$ consists of $\incx$, $\incy$, $\decx$, $\decy$, $\xzero$, and $\yzero$.

\smallskip
\noindent
{\bf Transition function.} The transition function $\delta_{\autom}$ of $\arpqm$ is a subset of $\states_{\autom} \times \Sigma_{\autom} \times \states_{\autom}$.
First, we shall explain the transitions of $\arpqm$ in a graphical manner. Given a state $\state \in \states$ recall the instruction of $\tca$ assigned to that state.
\begin{center}
\begin{minipage}{0.90\linewidth}
    
\begin{algorithmic}[1]
    \If{ $\cnt \texttt{ == } \mathtt{0}$}
    \State \texttt{$\cntp{}{1}$, move from $\state$ to $\statet$} 
    \Else  
    \State \texttt{$\cntp{}{\delt}$, move from $\state$ to  $\statef$} 
    \EndIf
\end{algorithmic}
\end{minipage}

\end{center}

Now, we take that instruction and with it we define the following transitions as depicted in the figure below:

\begin{figure}[H]
\centering
\includegraphics[width=0.85\linewidth]{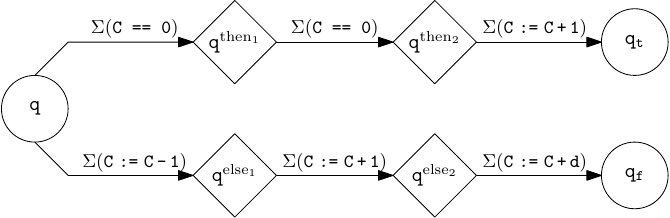}

\caption{ States other than $\state$, $\statef$, and $\statet$ are part of $\states_{aux}$ and are marked with diamonds in the figure. Label $\Sigma(\cnt \texttt{==} 0)$ evaluates to $\xzero$ or $\yzero$ depending whether $\cnt$ is $\cntl$ or $\cntr$; other labels evaluate to $\incx, \incy, \decx$, or $\decy$. The top branch of the figure corresponds to the ``\textbf{then}'' branch
\orange{of the instruction} assigned to the state $\state$ of $\tca$. \rebut{Note that $\Sigma(\cnt \texttt{==} 0)$ appears twice on the top branch. This redundancy is added to ensure an equal number of states on each branch toward establishing \cref{obs:threesteps}; it is by no means critical to the proof.}}
\end{figure}

The above figure encodes six triples from the set $\states_{\autom} \times \Sigma_{\autom} \times \states_{\autom}$ (for example, $\pair{\state,\; \Sigma(\mathtt{C == 0}),\; \state^{\mathrm{then}_1}}$ is one of them). To obtain the transition function $\delta_{\autom}$ of $\arpqm$, we take the union of the above-defined triples over the instruction set of $\tca$.

\medskip
\noindent
{\bf Definition of the query $\rpqm$. }
Let $A$ be any regular expression that defines the same regular language as $\arpqm$, and let $B$ be the regular expression resulting from concatenating $\xzero$, $\yzero$, and $A$. Then the Boolean RPQ $\rpqm$ is expressed as $\Exists{x,y}B(x,y)$.

\subsection{Correspondence between $\rpqm$ and $\tca$}
In this section we shall show the following lemma:

\begin{lemma}\label{lem:main-undecidable}
$\chase{\dbg, \rsg} \models \rpqm$ if and only if $\tca$ halts starting from $\pair{0,0,\state_{\mathtt{}0}}$.
\end{lemma}

To this end, we shall imagine a $B$-path corresponding to a match of $\rpqm$ to $\ginst$. Its two first steps have to be over symbols $\xzero$ and $\yzero$ which is possible only at \rebut{the} $\pair{0,0}$ coordinates of the grid $\ginst$. After this point, the query uses the automaton $\arpqm$ to define its behavior. Therefore, we will be discussing states and transitions of $\arpqm$ in the context of $\tca$. Imagine the automaton $\arpqm$ starting at coordinates $\pair{0,0}$ and ``walking'' over $\ginst$ and ``reading'' its binary predicates.
We say that the automaton $\arpqm$ at coordinates $\pair{x,y}$ and in state $\state$ is in \emph{configuration} $\pair{x,y,\state}$. To prove \cref{lem:main-undecidable}, and thus \cref{thm:main-undecidable}, it is enough to inductively use the following observation:

\begin{observation}\label{obs:threesteps}
For \rebut{all} natural numbers $x,x',y$, and $y'$ and every pair of states $\state, \state' \in \states$
the following two are equivalent:

\begin{itemize}
    \item Automaton $\arpqm$ transitions from configuration $\pair{x,y,\state}$ to $\pair{x',y',\state'}$ in three steps.
    \item TCA $\tca$ transitions \orange{from 
    $\pair{x,y,\state}$} to $\pair{x',y',\state'}$ in one step.
\end{itemize}
\end{observation}
\begin{proof}
\pagebreak

We shall inspect the case when the instruction from \cref{fig:instr-example} is assigned in $\tca$ to the state $\state$.  
\begin{figure}[H]
\centering
\begin{minipage}{0.90\linewidth}
\begin{algorithmic}[1]
    \If{ \texttt{$\cntr$ == $\mathtt{0}$}}
    \State $\cntp{y}{1}$\texttt{, move from $\state$ to $\mathtt{q_1}$}
    \Else  
    \State $\cntm{y}{1}$\texttt{, move from $\state$ to $\mathtt{q_2}$}
    \EndIf
\end{algorithmic}
\end{minipage}
\caption{Instruction assigned to state $\state$ of TCA $\tca$.
}\label{fig:instr-example}
\end{figure}
Then the part of $\arpqm$ responsible for transitioning out of $\state$ has the following form: 
\begin{figure}[H]
\centering
\includegraphics[width=0.85\linewidth]{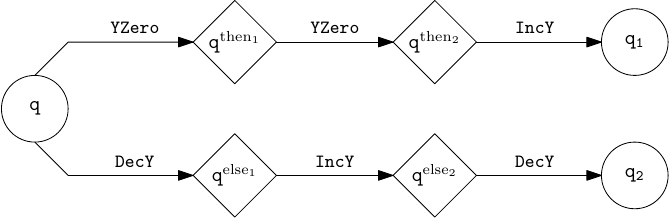}
\end{figure}

Assume $y = 0$. Then $\tca$ transitions to $\pair{\state_\mathtt{1}, x, 1}$. We argue that the automaton $\arpqm$ follows in three steps to the same configuration. First, note that the automaton cannot transition through the bottom branch of \cref{fig:instr-example} as it cannot follow a $\decy$-edge in $\ginst$ while being at zero $\ypred$-coordinate. However, $\arpqm$ can move through the top branch, transitioning to $\pair{\state_\mathtt{1}, x, 1}$.

Assume $y \neq 0$. Then $\tca$ transitions to $\pair{\state_\mathtt{2}, x, y \,{+}\, 1}$. Note that $\arpqm$ cannot follow the top branch (there is no $\yzero$ outgoing edge at $\pair{x,y}$ in $\ginst$, as $y > 0$). It can, however, follow the bottom one transitioning to $\pair{\state_\mathtt{2}, x, y - 1}$.
The other cases follow analogously.
\end{proof}

\section{Deciding RPQs over Sticky Rulesets}

This section is dedicated to the following theorem:

\begin{theorem}\label{thm:sticky-main}
RPQ entailment from sticky rulesets is decidable.
\end{theorem}

To simplify the intricate proof of the above, we restrict the query language to plain RPQs. However, it is straightforward to obtain the following as a consequence:

\begin{corollary}\label{thm:extended-sticky-main}
2RPQ entailment from sticky rulesets is decidable.
\end{corollary}

One can obtain \cref{thm:extended-sticky-main} by a simple reduction. Take a sticky ruleset $\rs$ over signature $\sig$, a database $\db$ and a 2RPQ $\rpq$. Keep $\db$ intact, add a single rule 
    ${\epred(x,y)}\to{{\epred'}(y,x)}$
to $\rs$ for every binary predicate $\epred \in \sig$, and replace every occurence of the inverted symbol $\inv{\epred}$ by \rebut{$\epred'$} in the regular expression of $\rpq$. 
It should be clear that entailment for the original problem coincides with that of the transformed problem. Finally, note that the transformation preserves stickiness of the ruleset.


The rest of the section is dedicated to the proof of \Cref{thm:sticky-main}. 
\textbf{\textit{Until the end of this section, we fix a database $\db$, and a sticky ruleset $\rs$.}}
We show decidability through two semi-decision procedures.

\noindent
\textbf{Caveat: no repeated variables in RPQs.} We emphasize that under our definition, RPQs of the form $A(x,x)$ and $\exists{x}.A(x,x)$ are not allowed, which is crucial for our proof. Queries with "variable joining" would require distinct techniques and would, arguably, better fit into the category of \emph{conjunctive} (2)RPQs.

\subsection{Recursive Enumerability}

This part exploits an easy reduction of RPQ entailment to a first-order entailment problem.

\begin{observation}\label{obs:datalog-stuff-in-RE}
Given a Boolean RPQ $Q = \Exists{x,y} A(x,y)$ one can write a Datalog ruleset $\rs_Q$ with one distinguished nullary predicate \textsc{Goal}, such that:
$$\database,\; \ruleset \models Q \quad\iff\quad \database,\; \rebut{\ruleset \,\cup\,} \rs_Q \models \textsc{Goal}.$$ 
\end{observation}
\begin{proof}
See supplementary material, \appwrap{\cref{app:obs:datalog-stuff-in-RE}}.
\end{proof}

As $\db \wedge \rs \wedge \rs_{\rpq}$ is an FO theory, we can invoke completeness of FOL to recursively enumerate consequences of $\db, \rs \cup \rs_{\rpq}$ and therefore semi-decide $\db, \rs \cup \rs_{\rpq} \models \textsc{Goal}$. Thus, we can semi-decide $\db, \rs \models \rpq$ as well.

\subsection{Co-Recursive Enumerability}
This part presents a greater challenge and necessitates a novel approach. We leverage the fact that, following a specific pre-processing of $\db$ and $\rs$, the non-entailment of RPQs is witnessed by a finite instance

\begin{definition}\label{def:finite-controllability}
Given a query language $\mathbb{L}$ and a ruleset $\rs$ we say that $\rs$ is {\em finitely $\mathbb{L}$-controllable} \iffi for every $Q \in \mathbb{L}$ and for every database $\db$ such that  $\db, \rs \not\models Q$ there exists a finite model $\mathcal{M}$ of $\db$ and $\rs$ s.t. $\mathcal{M} \not\models Q$.
\end{definition} 

\begin{lemma}\label{lem:sticky-co-semi-decision}
One can compute a database $\dbp$ and a ruleset $\rsp$ such that $\rsp$ is finitely RPQ-controllable \rebut{and such that for every Boolean RPQ $\rpq$ we have}  $\db, \rs \models \rpq \;\iff\; \dbp, \rsp \models \rpq$.
\end{lemma}

In view of this lemma, we can ``co-semidecide'' $\db, \rs \models \rpq$ (that is, semi-decide $\db, \rs \not\models \rpq$) 
by recursively enumerating all finite instances and terminating whenever a $\rpq$-countermodel of $\dbp, \rsp$ is found.

\subsection{Overview of the Proof of \cref{lem:sticky-co-semi-decision}}

Before we begin discussing the proof, let us introduce a specific class of rulesets:

\smallskip
\noindent
\textbf{Stellar rules.}
A {\em stellar rule} is a rule with at most one join variable, which also must be a frontier variable. A ruleset comprising only stellar rules is \emph{stellar}.

The ultimate goal is to construct $\rsp$ as a stellar ruleset. The reason behind this is as follows:

\begin{lemma}\label{lem:stellar-joinless-are-fc}
    Any ruleset consisting of stellar rules is finitely RPQ-controllable.
\end{lemma}
\noindent
We will first prove the above lemma, then we will show the construction of $\rsp$ and $\dbp$.

\subsection{Finite RPQ-Controllability}
\newcommand{\rss}{\mathcal{S}}
Assume $\rss$ is a stellar ruleset. To prove its finite RPQ-controllability, we present, given some database and a non-entailed RPQ, a construction of a finite countermodel. This involves two key parts: the "finite model" and the "counter" parts. We start with the first.

For any instance $\inst$ and two of its terms $t$, $t'$ we let $\inst[t,t' \mapsto t'']$ denote the structure obtained from $\inst$ by replacing each occurrence of $t$ or $t'$ by the fresh null $t''$\!\!. 

Preservation of modelhood under this kind of ``term merging'' is trivial for rules without join variables:

\begin{observation}\label{obs:modelhood-for-joinless}
    Let $\inst$ be an instance with terms $s,t$ and $\rho$ a joinless rule. \!\!Then $\inst \,{\models}\, \rho$ implies $\inst[s,t \,{\mapsto}\, u] \,{\models}\, \rho$. 
\end{observation}
\begin{proof}
    Note, as $\rho$ is joinless, identification of terms cannot create new active triggers.
\end{proof}

For stellar rules with a join variable, we define a kind of 1-type that indicates which terms are similar enough so they can be ``merged'' while preserving modelhood.

\smallskip
\noindent
\orange{\textbf{Stellar types.}}
A {\em stellar query} (SQ) is a CQ $S(y)$ such that $y$ is the only join variable of $S$ and each atom of $S$ has at most one occurrence of $y$.
Given a term $t$ of some instance $\inst$ the {\em stellar type} $\bigstar(t)$ of $t$ is the (up to equivalence) most specific\footnote{that is: minimal under CQ containment}
 SQ $S(y)$ such that $\inst \models S(t)$.

\begin{observation}\label{obs:modelhood-for-stellar}
Let $\inst$ be an instance with terms $s,t$ satisfying $\bigstar(s) = \bigstar(t)$ and $\rho$ a stellar rule with a single join variable.
Then $\inst \models \rho$ implies $\inst[s,t \mapsto u] \models \rho$. 
\end{observation}
\begin{proof}
Observe $\bigstar(s) = \bigstar(t) = \bigstar(u)$ and that the stellar types of other terms remain unchanged after identification ($\heartsuit$). Let $\alpha(\vx)$ be a head of $\rho$, with $\vx$ a tuple of its frontier variables. Note that as $\rho$ is stellar, $\vx$ contains its join-variable denoted by $x$. Let $H_v$ be the set of homomorphisms from a body of $\rho$ to $\inst$ such that $x$ is always mapped to $v$. Note, it can be determined whether every trigger in the set $\pair{\rho, h} \mid h \in H_v$ is satisfied based only on the stellar type of $v$. Therefore, from ($\heartsuit$), the satisfaction of $\alpha(\vx)$ is preserved.
\end{proof}

\begin{corollary}\label{cor:preserving-modelhood-for-s}
Let $\inst$ be an instance with terms $s,t$ satis\-fying $\bigstar(s) = \bigstar(t)$, and let $\rss$ be a stellar ruleset. 
Then $\inst \models \rss$ implies $\inst[s,t \mapsto u] \models \rss$.
\end{corollary}
\begin{proof}
From \cref{obs:modelhood-for-joinless}~and~\cref{obs:modelhood-for-stellar}.
\end{proof}


We now begin the ``counter'' part of the finite countermodel construction. Note its independence from $\rs$.
\newcommand{\regtyp}{\hspace{0.4mm}\updownarrow\hspace{-0.4mm}}

\newcommand{\skipit}[1]{}
\skipit{
\color{purple}
\noindent
\orange{\textbf{Regular types.}}
Given an DFA $\autom = \pair{\states, \Sigma, \delta, \state_0, \state_{\mathrm{fin}}}$, two of its states $\state, \state' \in \states$, an instance $\inst$, and one of its terms $t$, we write $\state\!\to\!\state'(t)$ (\,$\state\! \leftarrow\!\state'(t)$\,) \iffi there exists a term $t'$ of $\inst$ and a path from $t'$ to $t$ (from $t$ to $t'$) labelled with a word $w$ such that $\delta^*(\state, w) = \state'$. Define the {\em regular type} $\regtyp(t)$ of $t$ as the set of above-defined expressions holding for $t$ that $t$ satisfies w.r.t $\inst$ and $\autom$. 

\begin{lemma}\label{lem:regular-types-id}
    Given a Boolean RPQ $\rpq$ and a DFA $\autom$ defining it, an instance $\inst$, and two of its terms $t$ and $t'$ such that $\regtyp(t) = \regtyp(t')$. Then:
    $$\inst \models \rpq \quad\iff\quad \inst[t,t'\mapsto t''] \models \rpq.$$
\end{lemma}
\begin{proof}
We note that the regular types present in both $\inst$ and $\inst[t,t'\mapsto t'']$ are the same. Note, that for any instance $\jnst$ we have $\jnst \models \rpq$ \iffi there exists a term $s$ of $\jnst$ satisfying $\state_0 \to \state_{\mathrm{fin}} (s)$.
\end{proof}
}

\noindent
\orange{\textbf{Regular types.}}
Given a DFA $\autom = \pair{\states, \Sigma, \delta, \state_0, \state_{\mathrm{fin}}}$, an instance $\inst$, and one of its terms $t$, we define the {\em regular type} of $t$, denoted $\uparrow\!\!_\autom(t)$ as the set containing all expressions $\state\!\!\rightsquigarrow\!\!\state'$ for which $t$ has an outgoing $w$-path for some $w$ with $\delta^*(\state, w) = \state'$.

\begin{lemma}\label{lem:regular-types-id}
    Let $\rpq$ be a Boolean RPQ and $\autom$ the DFA $\autom$ defining it. Let $\inst$ be an instance and $t,t'$ two of its terms satisfying $\uparrow\!\!_\autom(t) = \uparrow\!\!_\autom(t)$. Then:
    $$\inst \models \rpq \quad\iff\quad \inst[t,t'\mapsto t''] \models \rpq.$$
\end{lemma}
\begin{proof}
We note that the regular types present in both $\inst$ and $\inst[t,t'\mapsto t'']$ are the same. On the other hand, for any instance $\jnst$ we have $\jnst \models \rpq$ \iffi there exists a term $s$ of $\jnst$ satisfying $\state_0\!\!\rightsquigarrow\!\!\state_{\mathrm{fin}} \in {\uparrow}\!_\autom(s)$.
\end{proof}


We are now ready to show that $\rss$ is finitely RPQ-controllable. Let $\db'$ be a database and $Q$ be a Boolean RPQ. Suppose $\chase{\db', \rss} \not\models Q$. Let $\inst$ be an instance obtained from $\chase{\db', \rss}$ by identifying all nulls that share the same regular and stellar types. With the support of \cref{cor:preserving-modelhood-for-s} and \cref{lem:regular-types-id}, given that there are only finitely many stellar and regular types for $Q$, we conclude that $\inst$ is finite, acts as a model of $\db'$ and $\rss$, and does not entail $Q$. Thus, \cref{lem:stellar-joinless-are-fc} is established.

\subsection{Construction of $\dbp$ and $\rsp$}
In this section, we construct $\dbp$ and $\rsp$ satisfying the two following lemmas. These, together with \cref{lem:stellar-joinless-are-fc}, establish \cref{lem:sticky-co-semi-decision}.

\begin{lemma}\label{lem:rsp-is-stellar-and-joinless}
    The ruleset $\rsp$ is stellar.
\end{lemma}

\begin{lemma}\label{lem:dbp-rsp-equiv-db-rs}
    For every Boolean query $Q$ it holds:
    $$\chase{\db, \rs}\models Q \iff \chase{\dbp, \rsp} \models Q.$$
\end{lemma}

In order to ensure the first of the two lemmas, we perform a couple of transformations on the initially fixed sticky ruleset $\rs$. As we progress, we maintain intermediate variants of the second lemma.

\subsubsection*{Rewriting-away non-stellar rules}
For this step, we heavily rely on the stickiness (and thus {\fus}ness) of $\rs$. The below relies on the idea that for \fus rulesets, rewriting can be also applied to bodies of rules -- yielding a new equivalent ruleset. This transformation preserves stickiness of the input ruleset and, importantly, it preserves Boolean RPQ entailment.

\newcommand{\rewop}{\mathtt{rew}}
\newcommand{\rrew}{\rew(\rs)}
\newcommand{\rhorew}{\rew(\rho, \rs)}
\begin{definition}\label{def:body-rewriting}
Given an existential rule $\rho \in \rs$ of the form:~$\cnarule{\alpha(\vx, \vy)}{\vz}{\beta(\vy, \vz)}$
let $\rhorew$ be the ruleset:
\begin{align*}
    &\Big\{\ \narule{\gamma(\vx', \vy)}{\vz}{\beta(\vy, \vz) \ \Big| \\[-1ex] 
    &\hspace{5ex}\Exists{\vx'}\gamma(\vx', \vy) \;\in\; \rew(\Exists{\vx} \alpha(\vx, \vy), \rs)}\ \Big\}.
\end{align*} 
Finally, let $\rrew= \rs \cup \bigcup_{\rho \in \rs}\; \rhorew.$
\end{definition}

\begin{lemma}\label{obs:rewriting-preserves-skicky}
The ruleset $\rrew$ is sticky.
\end{lemma}
\begin{proof}
See supplementary material, \appwrap{\cref{sec:rewriting-preserves-skicky}}.
\end{proof}

\begin{lemma}\label{lem:rrew-equiv}
For every Boolean RPQ $\rpq$ we have: $$\chase{\db, \rs} \models \rpq \iff \chase{\db, \rrew} \models \rpq.$$
\end{lemma}
\begin{proof}
See supplementary material, \appwrap{\cref{app:lem:rrew-equiv}}.
\end{proof}


We are now prepared to introduce the primary tool of the construction. The conceptual basis of this tool can be traced back to \citeauthor{journey-paper} (\citeyear{journey-paper}).

\begin{definition}\label{def:quick}
A ruleset $\rs'$ is {\em quick} \iffi for every instance $\inst$ and every atom $\beta$ of $\chase{\inst, \rs'}$ if all frontier terms of $\beta$ appear in $\adom{\inst}$ then $\beta \in \step{1}{\inst, \rs'}$.
\end{definition}

The technique of ``quickening'' \fus rulesets is quite versatile, allowing for the simplification of rulesets and the general streamlining of proof-related reasoning. 

\begin{lemma}\label{lem:frontier}
For any \fus ruleset $\rs$, $\rrew$ is quick.
\end{lemma}
\begin{proof}
Suppose $\beta(\vt, \vs) \in \chase{\inst, \rrew}$, where $\vt$ represents the frontier terms of $\beta$. Assume that $\vt$ appears in $\inst$, but $\beta$ does not.
Let $\rho$ be the rule that created $\beta$ during the chase and let it be of the following, general form:
$\;\narule{\alpha(\vx, \vy)}{\vz}{\beta(\vy, \vz)}.\;$
From the above, note that $\inst, \rrew \models \Exists{\vx}\alpha(\vx, \vt)$.

If $\rho \in \rs$ then there exists $\gamma(\vx', \vy)$ such that: 
$(\Exists{\vx'}\gamma(\vx', \vy)) \in \rew(\Exists{\vx} \alpha(\vx, \vy), \rs)$ and
$\inst \models \Exists{\vx'}\gamma(\vx', \vt)$. Therefore, from definition of $\rrew$, there exists a rule $\rho' \in \rrew$ (with $\gamma$ begin its body) such that $\rho'$ derives $\beta$ in $\step{1}{\inst, \rs'}$.

If $\rho \not\in \rs$ then let $\rho'$ be such that $\rho \in \rew(\rho', \rs)$ and $\rho' \in \rs$. As $\rho \in \rew(\rho', \rs)$  we know that $\rho'$ can be used as well during the chase to obtain $\beta$. Thus, we can use the above reasoning for $\rho'$ instead of $\rho$ to complete the proof of the lemma.
\end{proof}

\noindent
\textbf{Taking cores of rules.\;\;}
The ruleset we have ($\rrew$) is almost sufficient for constructing $\rsp$. Given our focus on sticky rulesets, we are particularly interested in variable joins appearing in rules. The rationale behind the following transformation step can be explained using the simple conjunctive query $\Phi(x) = \Exists{y, y'}\epred(x,y),\epred(x,y')$. The variable join on $x$ is unnecessary to express $\Phi$ as it is equivalent to its core $\Phi'(x) = \Exists{y} \epred(x,y)$. This motivates the following:

\begin{definition}\label{def:cored-rule}
Given an existential rule $\rho$ of the form $\alpha(\vx, \vy) \to \Exists{\vz} \beta(\vy, \vz)$, we define $core(\rho)$ 
as the existential rule $\alpha'(\vx', \vy) \to \Exists{\vz} \beta(\vy, \vz)$ where $\alpha'$ is a core of $\alpha$ with variables $\vy$ treated as constants.
\end{definition}

\begin{definition}\label{def:rnf}
Let $\rnf$ be $\set{\core(\rho) \mid \rho \in \rrew}$.
\end{definition}

As taking cores of bodies of rules produces an equivalent ruleset we have:

\begin{observation}\label{lem:rnf-good}
    $\chase{\db, \rnf} = \chase{\db, \rrew}$, and $\rnf$ is both sticky and quick.
\end{observation}
\newcommand{\rnfp}{\mathtt{cr}^{+}(\ruleset)}

\noindent
\textbf{Introducing redundant stellar rules.\;\;}
We are one step away from defining $\rsp$. Let $\rnfp$ be the ruleset obtained from $\rnf$ by augmenting it with a set of additional rules as follows: For every non-stellar rule $\rho$ in $\rnf$, let $\rnfp$ also contain the rule derived from $\rho$ by substituting all its join variables 
by one and the same fresh variable.
The following is straightforward:

\begin{observation}\label{lem:rnfp-good}
    $\chase{\db, \rnfp} = \chase{\db, \rrew}$, and $\rnfp$ is sticky and quick.
\end{observation}
\begin{proof}
    The first holds, as anytime a rule of $\rnfp \setminus \rnf$ is used to derive an atom, the original rule of $\rnf$ can be used instead. This argument also ensures the quickness of $\rnfp$. Finally, $\rnfp$ is sticky because the marking of variables for new rules of $\rnfp$ can be directly inherited from $\rnf$.
\end{proof}

Now we are ready to introduce $\rsp$. 

\begin{definition}
    Let $\rsp$ be obtained from $\rnfp$ by removing all rules with two or more join variables.
\end{definition}

The following lemma shows, that, 
as far as the derivation of binary atoms is concerned, 
$\rsp$ almost perfectly mimics $\rnfp$. The only limitation is that $\rsp$ might not be able to derive all binary atoms over database constants.


\begin{lemma}\label{lem:rsp-good-at-binary}
    $\chase{\db, \rsp}$ and $\chase{\db, \rnfp}$ when restricted to the binary atoms containing at least one 
    non-constant term 
    are equal.
\end{lemma}
\begin{proof}
    Consider $\chase{\db, \rnfp}$. 
    From the definition of $\rnfp$ we know that, whenever a rule containing more than one join variable produces an atom having exactly one and the same term on all its marked positions, then a rule from $\rnfp \setminus \rnf$ can be used to produce exactly the same atom. 
    Note this is also thanks to the specific version of the Skolem chase we use. Therefore, we will assume that rules of $\rnfp \setminus \rnf$ are used whenever possible.
    
    Consider all atoms in the chase $\chase{\db, \rnfp}$ that were necessarily generated by a rule having at least two distinct join variables -- that is, they couldn't have been created by any rule from $\rnfp \setminus \rnf$. Denote this set with $S$. Note that due to stickiness of $\rnfp$ (\cref{lem:rnfp-good}) atoms in $S$ have two distinct frontier terms on $\rnfp$-marked positions. Therefore, every atom that requires for its derivation an atom $\beta$ from $S$ has to contain both such terms from $\beta$ ($\clubsuit$).

    Let us categorize all binary atoms of $\chase{\db, \rnfp}$ into three distinct groups:
    \begin{enumerate}
        \item Atoms created by non-Datalog rules.
        \item Atoms created by Datalog rules and containing at least one term that is not a constant.
        \item Atoms containing two constants.
    \end{enumerate}

    To prove the lemma, it suffices to show that atoms of\\ the first and second kind are neither contained in $S$ ($\heartsuit$) nor does their derivation depend on atoms from $S$ ($\diamondsuit$).

    Atoms of the first kind trivially satisfy ($\heartsuit$): as $\rnfp$ is sticky, its non-Datalog rules can have at most one join variable. They also satisfy ($\diamondsuit$) since binary atoms created by non-Datalog rules cannot contain their birth term and two distinct frontier terms.

    For the second category of atoms, the argument is more involved. Let $\alpha(s,t)$ be an atom of the second kind and let $\rho$ be the rule that created $\alpha(s,t)$ in the chase. Let $h$ be the homomorphism witnessing this. Assume w.l.o.g. that $s$ is created in the chase no later than $t$ and that $t$ is not a constant.
    
    We shall argue ($\diamondsuit$) for $\alpha$ as follows. Assume that $t$ was created during the $(i{-}1)$th step of the chase, and let $\gamma(\vu, t)$ be the birth atom of $t$. Note that $\gamma(\vu, t)$ is the only atom containing $t$ in $\step{i-1}{\db, \rnfp}$, and as $\rnfp$ is quick, $\alpha(s,t)$ appears in $\step{i}{\db, \rnfp}$. Assume, towards a contradiction, that ($\diamondsuit$) does not hold for $\alpha(s,t)$. Therefore, there exist two distinct terms $u,v$ in $\step{i-1}{\db, \rnfp}$ that are contained in $\alpha(s,t)$, and these two terms are frontier terms of some atoms in $\step{i-1}{\db, \rnfp}$ (from $\clubsuit$), therefore both could not be $t$. As $\alpha(s,t)$ is a binary atom, we have a contradiction.

    Let us argue that ($\heartsuit$) holds for $\alpha(s, t)$. From above, we know that $\rho$ contains a $\gamma(\vx, y)$ atom in its body with $h(y) = t$, where $y$ is a frontier variable. Assume towards contradiction that $\rho$ has two distinct join variables. Denote them with $z$ and $w$. Note, as $\alpha(s,t)$ is binary, $z$ and $w$ map to $s$ and $t$ through $h$.  Without loss of generality assume $h(z) = s$ and $h(w) = t$. However, $\rho$ cannot have three distinct frontier variables -- each join variable should be a frontier variable from stickiness of $\rnfp$. Therefore $w = y$. We shall argue that $y$ cannot be a join variable. As $\rho$ is not stellar -- we assumed it has two distinct join variables $y$ and $z$ -- it is in $\rnf$. We shall show a contradiction with the fact that the body of $\rho$ is a core. Assume that $\gamma(\vu, t)$ is as in the argument above -- specifically it is the only atom containing $t$ just before creation of $\alpha(s,t)$. From this and the fact that $y$ is a join variable we note that $\gamma(\vx', y)$ is an atom of the body of $\rho$ and that $\vx \neq \vx'$. Note that the following cannot occur at the same time: 1) the body of $\rho$ contains both $\gamma(\vx, y)$ and $\gamma(\vx', y)$; 2) the body of $\rho$ has exactly two join variables; 3) the body of $\rho$ is a core. From this contradiction we get ($\heartsuit$) for $\alpha$. 
\end{proof}

Note that in the above, the only binary atoms that are missing are those which $\rnfp$ derives over database constants. This can be easily rectified:

\begin{definition}
Let $\dbp = \chase{\db, \rnfp}|_{\adom{\db}}$
\end{definition}

\cref{lem:sticky-co-semi-decision} requires that $\dbp$ is computable. Yet, as $\rnfp$ is sticky and therefore \fus, this is the case.

\begin{observation}\label{obs:dbp-is-computable}
$\dbp$ is computable.
\end{observation}
\begin{proof}
As $\rnfp$ is \fus, we rewrite every atomic query and ask if it holds for any tuple of constants of $\db$.
\end{proof}

\begin{corollary}
For every Boolean RPQ $Q$
    $\chase{\db, \rnfp} \models Q \iff \chase{\dbp, \rsp} \models Q.$
\end{corollary}
\begin{proof}
    Directly from \cref{lem:rsp-good-at-binary}, the fact that $\rsp \subseteq \rnfp$, and $\chase{\dbp, \rnfp} = \chase{\db, \rnfp}$.
\end{proof}

\noindent
From the above, \cref{lem:rrew-equiv} and \cref{lem:rnfp-good} we get:
$$\chase{\db, \rs} \models Q \iff \chase{\dbp, \rsp} \models Q$$
and therefore \cref{lem:dbp-rsp-equiv-db-rs}. Moreover, by the definition of $\rsp$ we get that $\rsp$ is stellar and thus we ensure \cref{lem:rsp-is-stellar-and-joinless}, concluding our overall argument.

\section{Undecidability and Stickiness}
In this section we show two seemingly harmless generalizations of the (decidable) case studied in the previous section. Importantly, both lead to undecidability and, given that both generalizations are rather slight, they highlight that the identified decidability result is not very robust.

\subsection{Generalizing RPQs}
Rather than having RPQs restricted to only use binary predicates, we may include higher-arity predicates, with the assumption that only the first two positions matter for forming paths. We refer to RPQs that permit such slightly extended regular expressions as {\em higher-arity regular path queries} (HRPQs). We obtain the following:

\begin{theorem}\label{thm:hrpq-undecidable-sticky}
    Boolean HRPQ entailment under sticky rulesets is undecidable.
\end{theorem}

In order to prove the theorem, one \rebut{can} use the proof of \cref{thm:main-undecidable} with a small tweak.
Consider the ruleset from \cref{def:undec-ruleset} with the last six rules tweaked:
\begin{align*}
\arule{\pnats(x,x')}{x''}{\pnats(x',x'')}\\
\arule{\pnats(x,x') \land \pnats(y, y')}{z}{\gridpred(x, y, z)}\\
\adrule{\gridpred(x,y,z)}{\xcoord(z, x)}\\
\adrule{\gridpred(x,y,z)}{\ycoord(z, y)}\\
\adrule{\varphi_\mathrm{right}(x,x',y,z,z')}{\incx(z,z',x,x',y)}\\
\adrule{\varphi_\mathrm{up}(x,y,y',z,z')}{\incy(z,z',x,y,y')}\\
\adrule{\incx(z,z',u,v,t)}{\decx(z',z,u,v,t)}\\
\adrule{\incy(z,z',u,v,t)}{\decy(z',z,u,v,t)}\\
\adrule{\xcoord(z, x) \land \zeropred(x)}{\xzero(z,z,x)}\\
\adrule{\ycoord(z, y) \land \zeropred(y)}{\yzero(z,z,y)}
\end{align*}
and note that it is sticky – there exists a trivial marking of positions for \orange{it.}
The rest of the proof of \cref{thm:main-undecidable} remains unchanged, observing that when we project $\incx, \incy, \decx, \decy, \xzero,$ and $\yzero$ to the first two positions, we obtain the previous ruleset.

\subsection{Generalizing stickiness}
Alternatively, instead of generalizing RPQs, one can consider ``slightly non-sticky'' rulesets.  Consider an extension of the above ruleset with the following projections. 
\begin{align*}
\adrule{\incx(z, z', u, v, t)}{\incx\predicate{Bin}(z, z')}\\
\adrule{\decx(z, z', u, v, t)}{\decx\predicate{Bin}(z, z')}\\
\adrule{\incy(z, z', u, v, t)}{\incy\predicate{Bin}(z, z')}\\
\adrule{\decy(z, z', u, v, t)}{\decy\predicate{Bin}(z, z')}\\
\adrule{\xzero(z, z, x)}{\xzero\predicate{Bin}(z, z)}\\
\adrule{\yzero(z, z, y)}{\yzero\predicate{Bin}(z, z)}
\end{align*}
Note that such expanded ruleset would as well admit undecidable RPQ entailment, and that the  projections are not allowing for further recursion. Therefore the above ruleset can be viewed as essentially sticky, just followed by one single projection step.

\section{Conclusion}
In this paper, we reviewed established existential rules fragments with decidable CQ answering, asking if the decidability carries over to RPQs.
We recalled that for the very comprehensive \fcs class of rulesets, the decidability for RPQs and even much more expressive query languages follows from recent results.
Thus focusing on the \fus class of rulesets, we established that they do not allow for decidable RPQ answering in general, due to the insight that -- unlike \fcs rulesets -- \fus rulesets allow for the creation of grid-like universal models, which then can be used as a ``two-counter state space'', in which accepting runs of two-counter machines take the shape of regular paths. On the other hand, we showed that 2RPQ answering over sticky rulesets is decidable thanks to the reducibility to a finitely RPQ-controllable querying problem. This decidability result is rather brittle and crucially depends on (1) the restriction of path expressions to binary predicates and (2)~the inability to freely project away variables in sticky rulesets. For this reason, a setting where RPQs are slightly liberalized leads to undecidability again.

There are several obvious questions left for future work:
\begin{itemize}
\item 
What is the precise complexity of 2RPQ answering over sticky rulesets? Recall that, as our decidability argument is based on finite controllability, the generic decision algorithm ensuing from that does not come with any immediate upper complexity bound.
\item 
Does the problem remain decidable for sticky rulesets when progressing to CRPQs or C2RPQs? 
We do not believe that a minor extension of our current proof would be sufficient to positively settle this question, as we are currently lacking methods of coping with variable joins in C(2)RPQs (under such circumstances, establishing finite controllability via stellar types fails). However, we think that some of the tools we used in this paper might come in handy when tackling that extended case.
\item 
What is the decidability status for other (non-\fcs) syntactically defined \fus fragments, such as \emph{sticky-join rulesets} \cite{DBLP:journals/ai/CaliGP12}? 
\item 
Is it possible and/or reasonable to establish a new class of rulesets, based on the rewritability of (C2)RPQs rather than CQs?
\end{itemize}

\section*{Acknowledgements}
Work supported by the European Research Council (ERC) Consolidator Grant 771779 (DeciGUT).

\ifcameraready
    \bibliographystyle{kr}
    \bibliography{20-bibliography}
    \newpage
    \phantom{a}
    \newpage
    \appendix
    \section{Proof of \cref{obs:datalog-stuff-in-RE}}\label{app:obs:datalog-stuff-in-RE}

Let $\autom$ be any DFA with its regular language being the same as the language of $A$. For every state $q_i$ of $\autom$ let $\qpred_i$ be a fresh unary relational symbol. Then the Datalog rule set $\rs_Q$ contains the following rules:
\begin{align*}
\adrule{}{\qpred_{0}(x)} \tag{for the starting state $q_0$}\\
\adrule{\qpred_{i}(x), \epred(x,y)}{\qpred_{j}(y)} \tag{if $\pair{q_i, \epred} \mapsto q_j$ is in $\autom$}\\
\adrule{\qpred_{f}(x)}{\textsc{Goal}}\tag{for the accepting state $q_f$}
\end{align*}

Essentially, chasing any instance with the above ruleset annotates every domain element with the states in which they can be reached by a walking automaton starting from some other domain element. \qed

\section{Proof of \cref{obs:rewriting-preserves-skicky}}\label{sec:rewriting-preserves-skicky}

In this section, we present a simplified version of the rewriting procedure from \cite{existential-rules-rewriting-procedure-montpelier} that is tailored to work for single-head rules. For the sake of completeness of the paper, we present the proof of the rewriting algorithm's soundness and completeness.

Given a CQ $Q$ with variables $\vx$ and an equivalence relation $\sim$ over $\vx$, we define $Q|_{\sim}$ as a CQ formed from $Q$ by inspecting each equivalence class $A$ of $\sim$ and replacing variables of $A$ in $Q$ with a fresh variable $x_A$; in case the equivalence class $A$ is a singleton, we keep the original variable. We say that $\sim$ is \emph{compatible} if each free variable of $Q$ is contained in a singleton equivalence class of $\sim$.

Given:
\begin{itemize}
    \item a CQ $Q(\vx)$ where $\vx$ is its tuple of free variables,
    \item a compatible equivalence relation $\sim$ over the variables of $Q$,
    \item a rule $\rho = \narule{\alpha(\vx, \vy)}{\vz}{\beta(\vy, \vz)}$,
    \item a isomorphism $f$ from the head of $\rho$ to $Q|_{\sim}$
\end{itemize}
we say that $\rho$ is {\em backward applicable} to $Q$ with respect to $f$ and $\sim$ if and only if

\begin{enumerate}
    \item $f(\beta(\vy, \vz))$ is the only atom of $Q|_{\sim}$ containing terms of $f(\vz)$,
    \item $f(\vz)$ consists of existentially quantified variables only.
\end{enumerate}

If, in addition to the above, the equivalence relation is minimal, meaning that for every $\sim' \subseteq \sim$, it is the case that $\rho$ is not backward applicable to $Q$ with respect to $f$ and $\sim'$, we write $\mathbf{back}(Q,\rho,f, \sim)$.

Now, assuming that $\mathbf{back}(Q,\rho,h, \sim)$ holds we denote with $\mathtt{rew}(Q, \rho, f, \sim)$ the following CQ:
$$Q|_{\sim} \setminus f(\beta(\vy, \vz)) \cup \alpha(\vx', f(\vy))$$
where $\vx'$ is a tuple of fresh variables.

\newcommand{\onerew}{\rewop_{\rs'}}
Given a \fus ruleset $\rs'$ and a UCQ $Q(\vvv)$ we define $\onerew(Q)$ as the following UCQ:
$$\onerew(Q) = \bigcup_{q \in Q}\set{\mathtt{rew}(q, \rho, f) \mid \Exists{\rho \in \rs', f} \mathbf{back}(q, \rho, f)}.$$
Moreover, by $\onerew^i$, we mean the i-fold composition of the $\onerew$ operator.

Given a \fus ruleset $\rs'$ and its rule $\rho = \narule{\alpha(\vx, \vy)}{\vz}{\beta(\vy, \vz)}$ let $\onerew(\rho, \rs')$ denote the following ruleset
\begin{align*}
\rs' \;\cup\; &\set{\narule{\gamma(\vx', \vy)}{\vz}{\beta(\vy, \vz)} \mid\\ &(\Exists{\vx'}\gamma(\vx', \vy)) \;\in\; \onerew(\Exists{\vx} \alpha(\vx, \vy))}.
\end{align*}

Let us first argue that $\onerew(\rho, \rs')$ is sticky if $\rs'$ is sticky.
\begin{lemma}\label{lem:one-step-rew-preserves-sticky}
    Given a sticky ruleset $\rs'$ and its rule $\rho$ we have that $\onerew(\rho, \rs')$ is sticky.
\end{lemma}
\begin{proof}
Take the body of $\rho$ and denote it with $\alpha(\vx, \vy)$ where $\vy$ is the tuple of its frontier variables. Let $M$ be the set of its variables that are either on a marked position or are join variables - it is the set of variables that have to appear on marked positions in the head of $\rho$. Let $\rho' = \narule{\gamma(\vx', \vy)}{\vz}{\beta(\vy, \vz)}$ be a rule created by the $\onerew(\rho, \rs')$ process, and let $\tau$ be the rule that was used to create $\rho'$ - therefore let $f$ and $\sim$ be such that $\mathbf{back}(\Exists{\vx} \alpha(\vx, \vy) ,\rho,f, \sim)$ holds. We shall argue that the set $M'$ of join variables and variables on marked positions of $\gamma(\vx', \vy)$ satisfies $M' \subseteq M$. Therefore $\rho'$ does not invalidate the original marking of $\rs'$. Let $Q$ denote $\Exists{\vx} \alpha(\vx, \vy)$ and $Q'$ denote $\Exists{\vx'}\gamma(\vx', \vy)$. Consider $Q|_{\sim}$. We shall argue that if a variable is in $M'$ - it is on a marked position in $Q'$ or is a join variable in $Q'$ - then it is on a marked position in $Q|{\sim}$. This implies $M' \subseteq M$. Note that $Q'$ is obtained from $Q|{\sim}$ by taking $\tau$ identifying which atom of $Q|{\sim}$ is going to be replaced with the body of $\tau$ using isomorphism $f$. Observe, however, that every join variable and every variable on a marked position added this way has to appear on a marked position in the head of $\tau$ therefore it appeared on a marked position in the atom that the head of $\tau$ maps to through homomorphism $f$.
\end{proof}

\begin{lemma}\label{lem:decidable-sticky-preservation-one-trigger-step}
Given a UCQ $Q(\vvv)$ and a \fus ruleset $\rs$, the following are equivalent for every instance $\inst$ and a tuple $\va$ of its elements:
\begin{enumerate}
	\item There exists a trigger $\tau$ in $\inst$ such that $\appl{\tau, \inst} \models Q(\va)$.
	\item $\inst \models \onerew(Q(\va))$.
\end{enumerate}
\end{lemma}
\begin{proof}
$(1 \Rightarrow 2).$\\
Let $\inst' = \appl{\inst, \tau}$, and let $q(\vvv)$ be a disjunct of $Q(\vvv)$ s.t. $\inst' \models q(\va)$. If $\inst \models q(\va)$ then $\inst \models \onerew(Q(\va))$ as $Q(\va) \subseteq \onerew(Q(\va)).$ Otherwise let $h$ be a homomorphism from $q(\va)$ to $\inst'$ and let $\gamma$ be the atom of $q(\va)$ that is mapped to the new atom of $\inst'$. Let $\sim$ be minimal equivalence relation over variables of $\gamma$ such that $f$ is an isomorphism from the head of $\rho$ to $\gamma|_{\sim}$, where $\rho$ is the rule in $\tau$. Finally, observe that $\inst \models \onerew(q(\va), \rho, h', \sim)$.

$(2 \Rightarrow 1).$\\
Let $q'(\va)$ be a CQ such that $q'(\va) \in \onerew(Q(\va))$ and $\inst \models q'(\va)$. Let $h_0$ be a homomorphism witnessing that $\inst \models q'(\va)$. If $q'(\va) \in Q(\va)$ then $\inst \models Q(\va)$. 

Take $\rho \in \rs$, a isomorphism $f$, equivalence relation $\sim$ over the variables of some disjunct $q$ of $Q$ such that $q' = \mathtt{rew}(q, \rho, f, \sim)$. Observe that there exists a homomorphism $h'$ such that $\appl{q'(\va), \pair{\rho, h'}}$ is isomorphic to $q'(\va) \cup q(\va)$ that is $\pair{\rho, h'}$ restores the atom of $q(\va)$ removed during $\mathtt{rew}(Q, \rho, h)$. From this we get that $q'(\va) \cup q(\va)$ maps to $\inst'$ and therefore $\inst' \models Q(\va).$
\end{proof}

\begin{corollary}
For every \fus ruleset and a UCQ $Q(\vvv)$ there exists a $k \in \nats$ such that for every instance $\inst$ and a tuple $\va$ of its terms we have:
$$\chase{\inst, \rs} \models Q(\va) \iff \inst \models \onerew^k(Q(\va)).$$
\end{corollary}
\begin{proof}
Follows from \cref{lem:decidable-sticky-preservation-one-trigger-step} and \cref{lem:bdd-is-fus}.
\end{proof}

From this we get, that applying $\onerew$ operator until the fixpoint is reached is a sound and complete rewriting procedure in the sense of definition \cref{def:rewriting}. Therefore we obtain \cref{obs:rewriting-preserves-skicky}.

\section{Proof of \cref{lem:rrew-equiv}}\label{app:lem:rrew-equiv}

\begin{lemma}\label{lem:rewriten-rs-equiv}
Both $\chase{\db, \rs}$ and $\chase{\db, \rrew}$ can be homomorphically mapped into each other.
\end{lemma}
\begin{proof}
As $\rs \subseteq \rrew$ it is enough to show that $\chase{\db, \rrew} \mapsto \chase{\db, \rs}$.

\newcommand{\chp}{\chase{\db, \rs}}
Take any instance $\inst$ such that there exists a homomorphism $h: \inst \to \chp$, and a trigger $\pi = \pair{\rho, f}$ in $\inst$ for a rule $\rho \in \rrew$. Let $\beta$ be the atom created from application of $\pi$ to $\inst$. We shall show that there exists a homomorphism \orange{$p':\inst \cup \beta \to \chp$}, then the lemma follows by a simple induction.

If $\rho \in \rs$, then $\pi' = \pair{\rho, h \circ f}$ is a trigger in $\chp$. Thus we can let $h'(\beta)$ to simply be the atom resulting from the application of $\pi'$ in $\chp$.

If $\rho \not\in \rs$. Let $\vt$ be the frontier terms of $\beta$, $\tau$ be a rule of $\rs$ such that $\rho \in \rew(\tau, \rs)$. Let $\alpha(\vx, \vy)$ and $\gamma(\vz, \vy)$ be the bodies respectively of $\rho$ and $\tau$ where $\vy$ is the tuple of their frontier variables. As $\inst \models \Exists{\vx} \alpha(\vx, \vt)$ from \orange{\cref{def:rewriting}~and~\cref{def:body-rewriting}} we know that $\inst, \rs \models \Exists{\vz}\gamma(\vz, \vt)$. From this we conclude that there is a trigger $\kappa = \pair{\tau, g}$ for $\tau$ in $\inst$ such that $\beta$ is also the resulting atom from application of $\kappa$ to $\inst$. As $\tau \in \rs$ we have reduced this case to the former.
\end{proof}

\else
    \bibliographystyle{abbrv}
    \bibliography{20-bibliography}
    \newpage
    \appendix
    
\fi

\end{document}